\documentclass[aps,prl,twocolumn,footinbib,superscriptaddress,longbibliography,amsmath,amssymb]{revtex4-1} 

\usepackage{graphicx}
\usepackage{subfigure}
\usepackage{epsfig} 
\usepackage{dcolumn}
\usepackage{bm}
\usepackage{times} 
\usepackage{color}

\usepackage[normalem]{ulem}

\begin{document}

\title{Signatures of the Higgs mode in transport through a normal-metal--superconductor
junction}

\author{Gaomin Tang}
\affiliation{Department of Physics, University of Basel, Klingelbergstrasse 82, CH-4056
Basel, Switzerland}
\author{Wolfgang Belzig}
\affiliation{Fachbereich Physik, Universit\"{a}t Konstanz, D-78457 Konstanz, Germany}
\author{Ulrich Z\"{u}licke}
\affiliation{School of Chemical and Physical Sciences and MacDiarmid Institute for
Advanced Materials and Nanotechnology, Victoria University of Wellington, P.O. Box 600,
Wellington 6140, New Zealand}
\author{Christoph Bruder}
\affiliation{Department of Physics, University of Basel, Klingelbergstrasse 82, CH-4056
Basel, Switzerland}

\bigskip

\begin{abstract}
  A superconductor subject to electromagnetic irradiation in the terahertz range can show
  amplitude oscillations of its order parameter. However, coupling this so-called Higgs
  mode to the charge current is notoriously difficult. We propose to achieve such a
  coupling in a particle-hole-asymmetric configuration using a DC-voltage-biased
  normal-metal--superconductor tunnel junction. Using the quasiclassical Green's function
  formalism, we demonstrate three characteristic signatures of the Higgs mode: (i) The AC
  charge current exhibits a pronounced resonant behavior and is maximal when the radiation
  frequency coincides with the order parameter. (ii) The AC charge current amplitude
  exhibits a characteristic nonmonotonic behavior with increasing voltage bias. (iii) At
  resonance for large voltage bias, the AC current vanishes inversely proportional to the
  bias. These signatures provide an electric detection scheme for the Higgs mode.
\end{abstract}

\maketitle

{\it Introduction.--}
Manipulating the superconducting (SC) state using tailored light pulses is currently
receiving a great deal of attention. Various fascinating phenomena have been reported,
including superconductivity enhancement \cite{Dayem-Wyatt66, Dayem-Wyatt67, Chang1977,
enhance18, enhance_cavity, enhance_CDW}, light-induced superconductivity \cite{Fausti189,
Mitrano2016, Jaksch19, Hart19}, the presence of chiral Majorana modes in chiral
superconductors \cite{Claassen19}, and the emergence of the Higgs mode \cite{Anderson63,
  Varma02, Varma15, Anderson15, Higgs_Raman_80, Higgs_Raman_81, Higgs_Raman_18,
  Littlewood_Varma_81, Littlewood_Varma_82, Higgs_THz_13, Higgs_THz_14, Kemper15,
  Higgs15_pseudospin, Rabi-Higgs18, Higgs19_rev, Higgs_dwave_18, Heikkila19, Krull2016, Efetov17,
  Nakamura19, Higgs_polariton, Puviani20, Higgs_proximity, Higgs_chiral, Higgs_amp,
  Schwarz2020_1, Wu20, Puviani20}.

The Higgs mode is a gapped collective excitation consisting of the oscillation of the
order parameter amplitude in a system with spontaneous symmetry breaking \cite{Anderson63,
Varma02, Varma15, Anderson15}. In a superconductor where the $U(1)$ symmetry is
spontaneously broken, the order parameter amplitude can oscillate when the system is coupled
to external gauge fields. The presence of a Higgs mode in superconductors with charge
density waves was first observed using the Raman-scattering technique
\cite{Higgs_Raman_80, Higgs_Raman_81, Higgs_Raman_18} and later theoretically
interpreted~\cite{Littlewood_Varma_81, Littlewood_Varma_82}. 
However, the excitation and detection of the Higgs mode in superconductors without charge
density waves became experimentally possible only in the last decade due to the
experimental advance of ultrafast low-energy terahertz (THz) spectroscopy. 
Clear nonlinear optical signatures indicating the presence of a Higgs mode have been
observed using the pump-probe technique in both $s$-wave \cite{Higgs_THz_13, Higgs_THz_14,
Kemper15, Higgs15_pseudospin, Higgs19_rev} and $d$-wave \cite{Higgs_dwave_18}
superconductors. 
Since the Higgs mode is a scalar excitation, it is expected to couple to the external
electromagnetic field in a nonlinear way. A linear coupling enabled by the presence of a
supercurrent was theoretically proposed \cite{Efetov17} and experimentally verified
\cite{Nakamura19}.
Very recently, it was theoretically predicted that a Higgs mode can be observed through
its effect in the time-dependent spin current in a ferromagnet-superconductor junction
\cite{Heikkila19}. 

\begin{figure}
	\centering
	\includegraphics[width=\columnwidth]{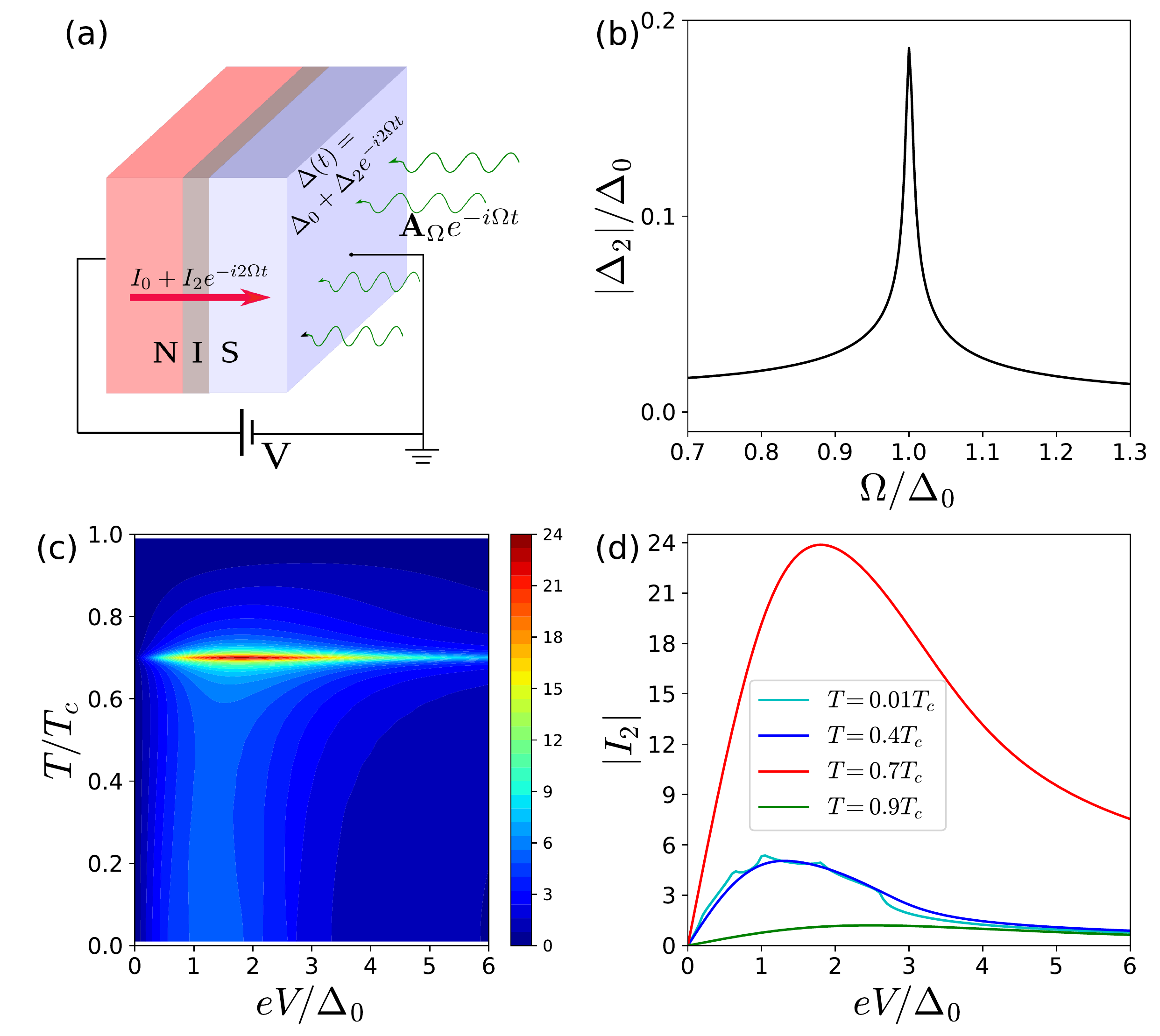} \\ 
  \caption{(a) DC-voltage-biased normal-metal--superconductor (NS) junction with a thin
    insulator (I) acting as a tunnel barrier. The superconductor is subject to
    electromagnetic irradiation described by a time-dependent vector potential
    $\mathbf{A}_\Omega e^{-i\Omega t}$, which generates a Higgs mode inside the
    superconductor. 
    (b) $|\Delta_2|/\Delta_0$ as a function of $\Omega/\Delta_0$. The parameters are
    chosen as $T=0.05T_c$, $\gamma=0.01\Delta_0|_{T=0}$, and $W_A=0.01\Delta_0|_{T=0}$.
    (c) AC charge current $|I_2|$ plotted against temperature $T$ and bias voltage $eV$
    scaled by the static SC gap amplitude at temperature $T$. The charge currents
    are in units of $G_t W_A/e$. The electromagnetic frequency is set to be $\Omega =
    \Delta_0|_{T=0.7T_c}$. (d) Line cuts of (c) at different temperatures.}
	\label{fig1}
\end{figure}

In this work, we consider a junction between a normal metal and an $s$-wave
superconductor [see Fig.~\ref{fig1}(a)]. THz electromagnetic irradiation can excite a
Higgs mode on the SC part of the junction.  To couple the Higgs mode to the charge
current, particle-hole symmetry has to be broken. Inducing a finite spin splitting in the
superconductor is one way to achieve this \cite{Heikkila19}.  As a potentially simpler
alternative, we propose to apply a DC voltage bias to a normal-metal--superconductor (NS)
junction.  We demonstrate that the Higgs mode will manifest itself in intriguing
properties of the AC charge current through the junction [see Fig.~\ref{fig1}(c)].

{\it Higgs mode in the superconductor.--}
We consider a superconductor subject to monochromatic electromagnetic irradiation with the
time-dependent vector potential $\mathbf{A}(t)=\mathbf{A}_{\Omega} e^{-i\Omega t}$ in the
Coulomb gauge.  
Due to the nonlinear coupling between the electromagnetic field and the Higgs mode, there
will be a second-harmonic correction to the static SC gap $\Delta_0$. Thus, the
time-dependent order parameter can be expressed as
\begin{equation}  \label{expand1} 
  \Delta(t) = \Delta_0 + \Delta_2 e^{-i2\Omega t} .
\end{equation}
If the irradiation frequency is close to $\Delta_0$, superconductivity can possibly be
enhanced~\cite{Chang1977, enhance18}, and a small correction to the static order parameter
can be incorporated in $\Delta_0$.

We employ the quasiclassical Green's function technique to study the dynamics of the
superconductor and the transport properties of the NS junction. The quasiclassical Green's
function $\check{g}_s(t,t')$ for a dirty superconductor with the diffusion constant $D$
fulfills the time-dependent Usadel equation~\citep{Belzig99, noneqSC}
\begin{equation}
  i\left\{ \check{\tau}_3\partial_t, \check{g}_s \right\} + \left[ \check{\Delta},
  \check{g}_s \right] - iD\nabla \left( \check{g}_s \circ \nabla \check{g}_s \right) =0,
  \label{Usadel}
\end{equation}
where the Green's function is written in Keldysh space and has the structure 
\begin{equation}
  \check{g}_s = 
  \begin{pmatrix}
    \hat{g}_s^r & \hat{g}_s^k \\ 0 & \hat{g}_s^a
  \end{pmatrix} .
\end{equation}
The order parameter has the form $\check{\Delta}=i\check{\tau}_2 \Delta$. The matrices
$\check{\tau}_{\alpha}$ with $\alpha=1,2,3$ are diagonal matrices with entries
$\hat{\tau}_{\alpha}$, that is $\check{\tau}_{\alpha}=\mathrm{diag}[\hat{\tau}_\alpha,
\hat{\tau}_\alpha]$, where $\hat{\tau}_\alpha$ are the Pauli matrices in Nambu space. In
Eq.~\eqref{Usadel}, the anti-commutator is defined as $\left\{ \check{\tau}_3\partial_t,
\check{g}_s \right\} = \check{\tau}_3 \partial_t \check{g}_s(t,t') + \partial_{t'}
\check{g}_s(t,t') \check{\tau}_3$, and $\nabla \cdot = \partial_\mathbf{r} \cdot -
ie[\check{\tau}_3 \mathbf{A}(t), \cdot]$. The convolution operation between two objects
$f$ and $g$ is defined as $(f \circ g)(t,t') = \int dt_1 f(t,t_1) g(t_1,t')$. The Green's
function obeys the normalization condition $\check{g}_s \circ \check{g}_s = 1$.

Since the Higgs mode couples to the electromagnetic field nonlinearly, to leading order
there is a second-harmonic correction $\check{g}_{2}$ to the stationary Green's function
$\check{g}_{0}$,
\begin{equation} \label{expand2} 
  \check{g}_s(t,t') = \check{g}_{0}(t-t') + \check{g}_2 (t, t').
\end{equation}
The Fourier transforms of $\check{g}_{0}$ and $\check{g}_{2}$ are, respectively, defined
as
\begin{align}
  \check{g}_{0}(t-t') &= \int \frac{d\epsilon}{2\pi} e^{-i\epsilon(t-t')}
  \check{g}_{0}(\epsilon), \notag \\
  \check{g}_2(t, t') &= \int \frac{d\epsilon}{2\pi} e^{-i\epsilon_+ t + i\epsilon_- t'}
  \check{g}_2(\epsilon_+, \epsilon_-) ,
\end{align}
with $\epsilon_\pm = \epsilon\pm \Omega$. The normalization condition leads to
$\check{g}_2(\epsilon_+, \epsilon_-)\check{g}_{0}(\epsilon_-)
+\check{g}_{0}(\epsilon_+)\check{g}_2(\epsilon_+, \epsilon_-)=0$. 

For the stationary quasiclassical Green's functions, we have~\cite{noneqSC}
\begin{align}
  & \hat{g}^{r(a)}_{0}(\epsilon) = g^{r(a)}_{0}(\epsilon)\hat{\tau}_3 + i\hat{\tau}_2
  f^{r(a)}_{0}(\epsilon) , \\
  & \hat{g}^{k}_{0}(\epsilon) = \left[\hat{g}^{r}_{0}(\epsilon) -
  \hat{g}^{a}_{0}(\epsilon)\right] \tanh\left( \beta\epsilon/2\right) ,
\end{align} 
where $\beta = 1/(k_B T)$, and 
\begin{equation}
  g_{0}^{r(a)}=f_{0}^{r(a)}\epsilon/\Delta_0 = \epsilon/s_0^{r(a)} ,
\end{equation}
with $s_0^{r(a)}=i\sqrt{\Delta_0^2-(\epsilon\pm i\gamma)^2}$. Here, $\gamma$ is the
phenomenological Dynes broadening parameter. Inserting Eqs.~\eqref{expand1} and
\eqref{expand2} into the Usadel equation \eqref{Usadel} leads to the retarded and advanced
components of the nonstationary term~\cite{Heikkila19}, 
\begin{equation} \label{g_2}
  \hat{g}^{r(a)}_2(\epsilon_+,\epsilon_-) = \hat{g}^{r(a)}_{V}(\epsilon_+,\epsilon_-) +
  \hat{g}^{r(a)}_{H}(\epsilon_+,\epsilon_-) .
\end{equation}
The first term $\hat{g}_{V}$ is due to the direct second-order coupling to the vector
potential and reads
\begin{equation}
  \hat{g}^{r(a)}_{V}(\epsilon_+,\epsilon_-) = iW_A \left[\frac{ \hat{\bar{g}}_{0}(\epsilon)
  - \hat{g}_{0}(\epsilon_+) \hat{\bar{g}}_{0}(\epsilon) \hat{g}_{0}(\epsilon_-) }{s_+ + s_-}
  \right]^{r(a)} ,
  \label{gv}
\end{equation}
with $W_A=D e^2|\mathbf{A}_\Omega|^2$ and $\hat{\bar{g}}_{0}= \hat{\tau}_3 \hat{g}_{0}
\hat{\tau}_3$. 
The second term $\hat{g}_{H}$ describes the effect of the Higgs mode, 
\begin{equation}
\hat{g}^{r(a)}_{H}(\epsilon_+,\epsilon_-) = i\Delta_2\left[\frac{ \hat{\tau}_2 -
\hat{g}_{0}(\epsilon_+) \hat{\tau}_2 \hat{g}_{0}(\epsilon_-) }{s_+ + s_-} \right]^{r(a)}.
\end{equation}
Both here and in (\ref{gv}), we have used the notation
$s_+^{r(a)}=i\sqrt{\Delta_0^2-(\epsilon_+\pm i\gamma)^2}$ and
$s_-^{r(a)}=i\sqrt{\Delta_0^2-(\epsilon_-\pm i\gamma)^2}$. 

An expression for the oscillating part $\Delta_2$ of the SC gap can be obtained from the
gap equation $\Delta(t)=-i\lambda \mathrm{Tr}[\hat{\tau}_2 \hat{g}^k_s(t)]$, where
$\lambda$ is the pairing interaction. The details of the derivation can be found in the
Supplemental Material~\cite{SM}. 
The static order parameter $\Delta_0$ at temperature $T$ (denoted by $\Delta_0|_T$) can be
well fitted by the interpolation formula
\begin{equation} \label{Delta_0}
  \Delta_0|_T = \Delta_0|_{T=0} \tanh(1.74 \sqrt{T_c/T-1}) ,
\end{equation}
where $T_c$ is the critical temperature.
To simplify the notation, we will omit the variable $T$ in $\Delta_0|_T$ at nonzero
temperature.

The Dynes broadening parameter and irradiation intensity are fixed as $\gamma =
0.01\Delta_0|_{T=0}$ and $W_A=0.01\Delta_0|_{T=0}$ in this work. In Fig.~\ref{fig1}(b), we
plot the dependence of $\Delta_2$ on the irradiation frequency $\Omega$ at temperature
$T=0.05T_c$. A cusp appears at $\Omega=\Delta_0$ (resonant condition) where the
oscillation amplitude of the order parameter $|\Delta_2|$ is maximal. 
The resonant behavior is a signature of the Higgs mode and can be identified in nonlinear
optical response using the pump-probe technique~\citep{Higgs_THz_13, Higgs_THz_14,
Kemper15, Higgs15_pseudospin, Higgs19_rev}.

{\it Normal-metal--superconductor junction.--}
In the following, we study the transport properties of a DC-voltage-biased NS junction, in
which the SC side is subject to electromagnetic irradiation described by a vector
potential ${\bf A}(t)=\mathbf{A}_{\Omega} e^{-i\Omega t}$ [see Fig.~\ref{fig1}(a)]. The
THz electromagnetic field is assumed to exist only on the SC side to avoid possible
photon-assisted tunneling processes, and its wave vector is parallel to the transport
direction of the junction.
The thickness of the superconductor is assumed to be smaller than or comparable to the SC
coherence length, so that the order parameter can be treated as homogeneous. For
simplicity, we consider a tunnel junction, which is characterized by its conductance
$G_t$. The retarded, advanced and Keldysh components of the Green's function for the
normal metal are expressed as
\begin{equation}
  \hat{g}^r_n = -\hat{g}^a_n = \hat{\tau}_3 ,
\end{equation}
and
\begin{equation}
  \hat{g}^k_n(\epsilon) = 2 \ \mathrm{diag} \Big[\tanh \big(\frac{\epsilon-eV}{2k_BT} \big),
  \ -\tanh \big(\frac{\epsilon+eV}{2k_BT} \big) \Big] ,
\end{equation}
where $V$ is the external voltage bias.

Since the leading perturbation due to the Higgs mode is a second harmonic in $\Omega$, the
electric current $I(t)$ can be decomposed in a DC component $I_0$ and an AC component
$I_2$ with $I(t) = I_0 + I_2 e^{-i2\Omega t}$. For a tunnel junction, the AC component of
the particle current to the first order in the tunnel conductance is
\begin{equation} \label{I2}
  I_2 = \frac{G_t}{8e} \int d\epsilon \ \mathrm{Tr} ( \hat{\tau}_3 \hat{X} ) ,
\end{equation}
with 
\begin{equation}
  \hat{X} = \hat{g}^k_n(\epsilon_+)\hat{g}^a_2(\epsilon_+,\epsilon_-)
  -\hat{g}^r_2(\epsilon_+,\epsilon_-) \hat{g}^k_n(\epsilon_-) . 
\end{equation}
Equation~\eqref{I2} implies that $I_2|_{eV=0}=0$, since the situation is particle-hole
symmetric in the absence of a DC voltage bias. A finite voltage bias can break the
symmetry so that the Higgs mode can couple to the charge current. This is in contrast to
Ref.~\citep{Heikkila19}, where particle-hole asymmetry is due to the exchange field in the
superconductor induced by an external magnetic field. A detailed derivation of $I_2$ using
circuit theory~\cite{circuit99, Bruder02, circuit11} can be found in the Supplemental
Material~\cite{SM}. 

\begin{figure}
\centering
\includegraphics[width=\columnwidth]{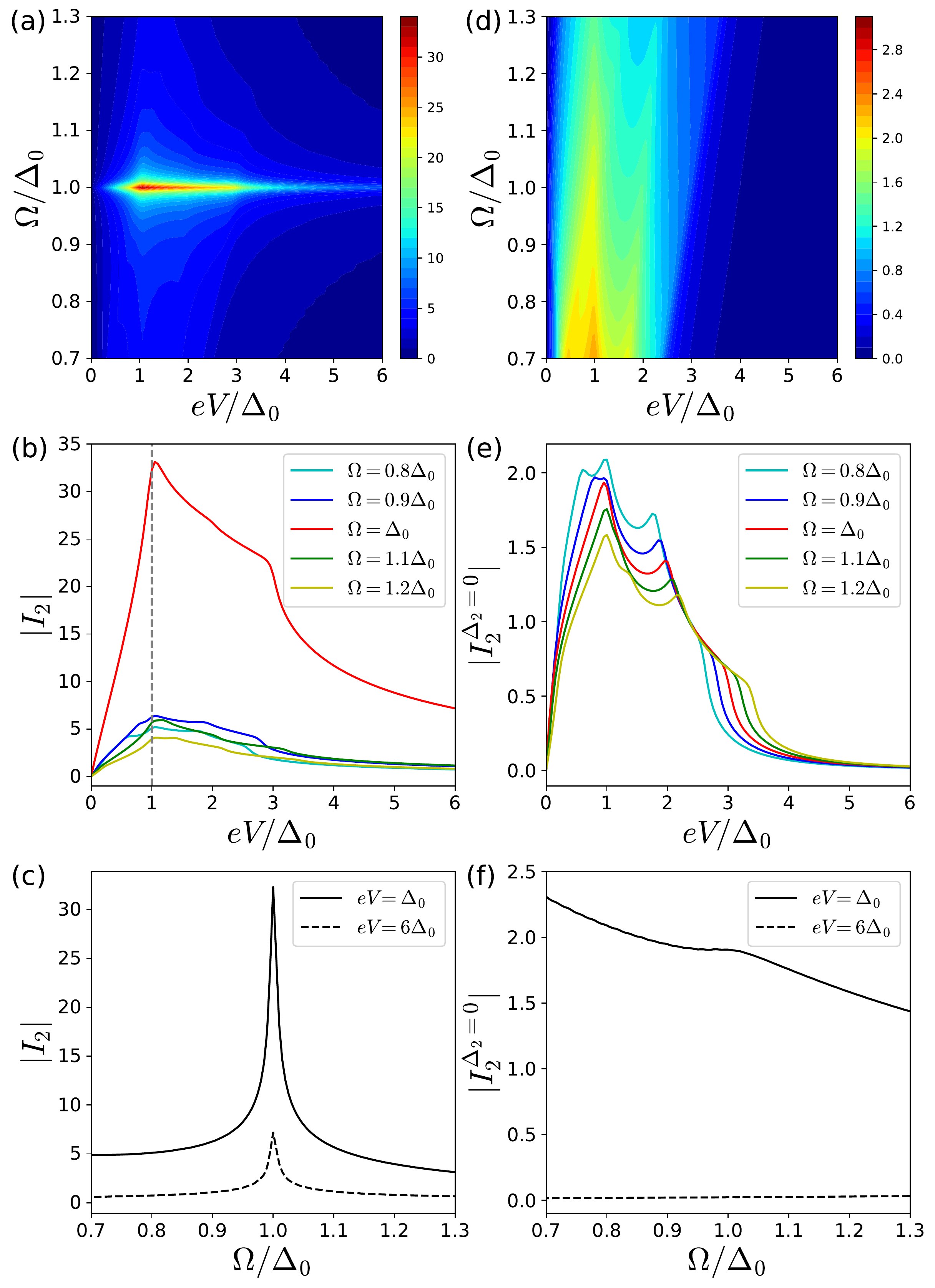} 
\caption{(a) AC charge current $|I_2|$ plotted against voltage bias $eV$ and irradiation
  frequency $\Omega$. Panels (b) and (c) are line cuts of (a) at different $\Omega$ and
  $eV$, respectively. (d) AC charge current $|I_2^{\Delta_2=0}|$ in the absence of the
  Higgs mode, i.e. by artificially setting $\Delta_2 = 0$. Panels (e) and (f) are line
  cuts of (d) at different $\Omega$ and $eV$, respectively. The charge currents are in
  units of $G_t W_A/e$. The temperature is $T=0.05T_c$, the other parameters are the same as
  in Fig.~\ref{fig1}(b).}
\label{fig2}
\end{figure}

{\it Numerical results.--}
The gap oscillation amplitude $\Delta_2$ is a complex number. This also applies to the AC
charge current $I_2$, and we focus on discussing the AC charge current magnitude $|I_2|$.
Figures~\ref{fig2}(a), (b) and (c) show the dependence of $|I_2|$ on the voltage bias $eV$
and the electromagnetic frequency $\Omega$, where panels (b) and (c) are line cuts of
panel (a) for different values of $\Omega$ and $eV$, respectively. To illustrate the
influence of the Higgs mode on transport, we artificially set $\Delta_2=0$ in
$\check{g}_2$, that is, $\hat{g}_2 = \hat{g}_{V}$ from Eq.~\eqref{g_2}, and calculate the
corresponding AC current amplitude $|I_2^{\Delta_2=0}|$ for comparison [see
Figs.~\ref{fig2}(d), (e) and (f)]. The charge currents are shown in units of $G_t W_A/e$.

As can be seen from Fig.~\ref{fig2}, the AC charge current amplitude $|I_2|$ at a fixed
voltage bias shows a pronounced resonant behavior as a function of frequency at
$\Omega=\Delta_0$, which is not present for $|I_2^{\Delta_2=0}|$. This can be easily
observed by comparing Fig.~\ref{fig2}(c) and (f) at $eV=\Delta_0$. Also, $|I_2|$ is much
larger than $|I_2^{\Delta_2=0}|$ at resonance, since the SC gap oscillation amplitude
$|\Delta_2|$ achieves its maximum and is much larger than $W_A$ as can be seen from
Fig.~\ref{fig1}(b). Thus, the Higgs mode dominates the AC charge current at resonance. In
Fig.~S1 in the Supplemental Material~\cite{SM}, we compare the contributions from
$\hat{g}_V$ and $\hat{g}_H$ in the AC charge current amplitude and find that the
contribution from $\hat{g}_V$ can be even larger than that from $\hat{g}_H$ away from
resonance. The resonant behavior of $|I_2|$ can serve as a signature of the presence of
the Higgs mode in the superconductor. A similar observation has been recently reported for
spin currents driven by the Higgs mode~\cite{Heikkila19}.

Figure~\ref{fig2}(b) shows that $|I_2|$ exhibits an interesting non-monotonic behavior: it
increases with increasing voltage bias up to around $eV=\Delta_0$, then starts to
decrease. At resonance $\Omega=\Delta_0$, the AC charge current shows a peak around
$eV=\Delta_0$.
This is explained as follows. The Higgs mode can be interpreted as coherent depairing and
pairing of Cooper pairs. The characteristic frequency of the Higgs mode is $2\Delta_0$,
thus the split quasiparticles (or depaired Cooper pairs) will appear around the SC band
edge. Since it is the split quasiparticles that contribute to the AC charge current at
resonance, the AC charge current reaches a maximum if the DC voltage bias aligns with the
SC band edge at low temperatures. There is also a kink at $eV=3\Delta_0$ that can be
interpreted as a side band due to the Higgs mode that modulates the quasiparticle density. 

Away from resonance, we can observe some other special points in Fig.~\ref{fig2}(b). These
points are due to the excitation from the electromagnetic irradiation with maxima at
$eV=\Delta_0$, $-\Delta_0+2\Omega$, $\Delta_0+\Omega$ and a kink at $\Delta_0+2\Omega$ as
can be observed from Fig.~\ref{fig2}(e). At higher temperatures, the non-monotonic
behavior of $|I_2|$ still survives, while the maximum point shifts to higher voltages with
$eV > \Delta_0$ due to thermal excitations (see Fig.~S2 in the Supplemental
Material~\cite{SM} for $T=0.7T_c$).

Under a large bias, the AC charge current $I_2$ shows characteristic features: even at
$eV=6\Delta_0$, a resonant behavior as a function of $\Omega$ can be observed in
Fig.~\ref{fig2}(c). In contrast, if the Higgs mode is not taken into account, the AC
charge current $I_2^{\Delta_2=0}$ vanishes at large voltage bias even for
$\Omega=\Delta_0$ as shown in Fig.~\ref{fig2}(e). At resonance and with $eV\gg \Delta_0$,
an analytical estimate given in the Supplemental Material \cite{SM} shows that the AC
current is inversely proportional to the voltage bias, i.e., $I_2 \propto \Delta_2/(eV)$.
Thus a finite AC charge current can persist up to relatively large voltage biases. This
characteristic behavior of the AC charge current at large bias near resonance can serve as
an indicator for the presence of the Higgs mode in the superconductor.


Apart from the possibility to tune the irradiation frequency continuously in experiment
\cite{THz17,THz20}, 
one can alternatively tune the Higgs mode to be at or away from the resonant condition by
changing the system temperature \cite{Higgs_THz_14}. In Fig.~\ref{fig1}(c), we show the AC
charge current magnitude as a function of voltage bias and temperature. The
electromagnetic frequency is fixed at $\Omega = \Delta_0|_{T=0.7T_c} \approx
0.8\Delta_0|_{T=0}$, which means that the resonance is achieved by tuning the temperature
to be at $T=0.7T_c$. This leads to a prominent resonant behavior as shown in
Fig.~\ref{fig1}(c). 

Figure~\ref{fig1}(d) also shows that $|I_2|$ exhibits a nonmonotonic behavior as a
function of voltage bias. At a low temperature with $T=0.01T_c$, $|I_2|$ reaches its
maximum around $eV=\Delta_0$. 
It also has maxima at both
$0.6\Delta_0$ ($\approx-\Delta_0+2\Omega$) and
$1.8\Delta_0$ ($\approx\Delta_0+\Omega$) and a kink at
$2.6\Delta_0$ ($\approx\Delta_0+2\Omega$) due to photon-assisted
transport processes. Due to thermal excitations, the voltage bias at which $I_2$ reaches
its maximum increases with increasing temperature. Once the temperature reaches the SC
critical temperature, both the static SC gap amplitude and its oscillation vanish, and so
does the AC charge current. Similarly to Fig.~\ref{fig2}(b), we observe a slow decay
$\propto \Delta_2/(eV)$ of the AC charge current at large bias near resonance. 

Since $\Delta_2$ is proportional to $W_A$ \cite{SM}, the next-order correction to
  the time-dependent order parameter $\Delta(t)$ in Eq.~\eqref{expand1} is proportional to
  $W_A^2$.  Higher-order corrections to $\Delta(t)$ can be ignored if
  $W_A/\Delta_0\ll 1$
  is satisfied. Given the smallness of $W_A/(\Delta_0|_{T=0})$, which is $0.01$ in this
  work, the condition $W_A/\Delta_0 \ll 1$ can be satisfied if the temperature is not too close
  to $T_c$.

The AC current $I_2$ is given in units of $G_t W_A/e$ where
  $G_t$ is the tunnel conductance of the junction and the value chosen
  for the coupling parameter $W_A$ is $W_A=0.01\Delta_0|_{T=0}$. For example,
  for the superconductor NbN, $\Delta_0|_{T=0} \approx
  2.6\,$meV~\cite{Nakamura19}, so that the values of $I_2$ shown in
  the figures are approximately in units of $0.026\,{\rm mV}\cdot
  G_t$. 

{\it Discussion.--}
The physical picture that emerges is as follows:
the pairing/depairing processes associated with the order-parameter
oscillation (Higgs mode) will create particles and holes each of which
contribute to the AC tunnel current. At zero voltage bias, their
respective contributions cancel. A finite voltage bias leads to
particle-hole asymmetry that results in a finite AC current that
exhibits signatures of the Higgs mode. On the one hand, this AC charge
current may be measured directly. Alternatively, the NS junction can
be experimentally manufactured as an antenna, where the
electromagnetic irradiation magnitude and frequency due to the AC
charge current can be detected. Even though a tunnel junction is
studied here, we expect all of the discussed features to be present
for a highly transparent junction as well.
Note that it is well established that nonequilibrium achieved by
voltage biasing can affect the AC response of a superconductor (see,
e.g.,~\cite{AC10}). However, our work considers the inverse
situation that the AC response of a superconductor (the Higgs mode)
can affect the transport properties of a voltage-biased junction.

The order-parameter oscillation will slowly decay in a power-law way if the
electromagnetic irradiation is switched off, and the same applies to the AC charge
current. Here, we only treat the steady-state situation with a constant electromagnetic
irradiation. 

At low temperatures $T \ll \Delta_0$, the conductance of the NS-junction that we
  consider is exponentially suppressed. Thus, for the results presented in
  Fig.~\ref{fig2}, where the temperature is chosen as $T=0.05 T_C$, we do not expect Joule
  heating to be a severe problem, and the predicted effects will be unchanged. For the results
  presented in Fig.~\ref{fig1}(c) and (d), if Joule heating exists, the resonant behavior
  will occur at a lower temperature. However, we expect all the qualitative features to
  remain.

In the work presented above, the wave vector of the electromagnetic field is assumed to be
parallel to the transport direction of the junction. If the wave vector has a finite
orthogonal component, the supercurrent, which can be induced by the Andreev reflection
processes, will mediate a linear coupling between the Higgs mode and the electromagnetic
field~\cite{Efetov17, Nakamura19, Higgs_polariton,Puviani20}. In this case, the
time-dependent order parameter can be written as $\Delta(t)=\Delta_0+\Delta_1 e^{-i\Omega
t} +\Delta_2 e^{-i2\Omega t}$ with an additional term of $\Delta_1$ compared to
Eq.~\eqref{expand1}. $\Delta_1$ and $\Delta_2$ exhibit resonances at $\Omega=2\Delta_0$
and $\Omega=\Delta_0$, respectively. AC charge transport in an NS junction in the presence
of a linear coupling between the Higgs mode and the electromagnetic irradiation will be
investigated in the future. The generalization to an NS junction with unconventional
superconductors \cite{Higgs_dwave_18,Schwarz2020_1, Wu20} may also be interesting. 

{\it Conclusion.--}
We have studied the AC transport properties of a DC-voltage-biased NS tunnel junction
taking into account the Higgs mode in the superconductor. The pronounced resonant
behavior, the characteristic nonmonotonic behavior of the AC charge current with
increasing bias and its slow decay inversely proportional to the bias can serve as
signatures of the presence of the Higgs mode.

Our results could be applied to design more complex superconducting devices featuring the
Higgs dynamics in coupled junctions of superconducting islands. Furthermore, it will be
interesting to combine Josephson effects with the Higgs mode.

\begin{acknowledgments}
{\it Acknowledgments.--} 
G.T. and C.B. acknowledge financial support from the Swiss National Science Foundation
(SNSF) and the NCCR Quantum Science and Technology.
\end{acknowledgments}

\bibliography{bib_higgs_NS}{}

\newpage
$\,$
\newpage

\begin{center}
\large\bf Supplemental Material
\end{center}

\section{Usadel equation and oscillating part of superconducting order parameter}
\label{appendixA}
We consider a monochromatic electromagnetic irradiation with the time-dependent vector
potential $\mathbf{A}(t)=\mathbf{A}_{\Omega} e^{-i\Omega t}$. The Higgs mode manifests
itself in a time-dependent order parameter,
\begin{equation} \label{expand1_A} 
\Delta (t) = \Delta_0 + \Delta_2 e^{-i2\Omega t}.
\end{equation}
The quasiclassical Green's function $\check{g}_s(t,t')$ for a dirty superconductor with
the diffusion constant $D$ fulfills the time-dependent Usadel equation \cite{noneqSC,
Belzig99}, 
\begin{equation} \label{Usadel_A}
i\left\{ \check{\tau}_3\partial_t, \check{g}_s \right\} + \left[ \check{\Delta},
\check{g}_s \right] - iD\nabla \left( \check{g}_s \circ \nabla \check{g}_s \right) =0.
\end{equation}
Here, the anti-commutator is defined as 
\begin{equation}
\left\{ \check{\tau}_3\partial_t, \check{g}_s \right\} = \check{\tau}_3 \partial_t
\check{g}_s(t,t') + \partial_{t'} \check{g}_s(t,t') \check{\tau}_3 ,
\end{equation}
and $\nabla \cdot = \partial_\mathbf{r} \cdot - ie[\check{\tau}_3 \mathbf{A}(t), \cdot]$.
The convolution operation between two objects $f$ and $g$ is defined as 
\begin{equation}
(f \circ g)(t,t') = \int dt_1 f(t,t_1) g(t_1,t').
\end{equation}
The quasiclassical Green's function is written in Keldysh space and has the structure
\begin{equation}
\check{g}_s = \begin{pmatrix}
\hat{g}_s^r & \hat{g}_s^k \\  &  \hat{g}_s^a
\end{pmatrix} .
\end{equation}
The superconducting (SC) order parameter has the form $\check{\Delta}=i\check{\tau}_2
\Delta$. The matrices $\check{\tau}_{\alpha}$ with $\alpha=1,2,3$ are diagonal matrices
with entries $\hat{\tau}_{\alpha}$, that is
$\check{\tau}_{\alpha}=\mathrm{diag}[\hat{\tau}_\alpha, \hat{\tau}_\alpha]$, where
$\hat{\tau}_\alpha$ are the Pauli matrices in the Nambu space. The quasiclassical Green's
function obeys the normalization condition $\check{g}_s \circ \check{g}_s = 1$. We will
assume all quantities to be uniform within the SC or normal side of the junction. To
leading order, the Green's function and order parameter can be expressed as
\begin{equation} \label{expand2_A} 
\check{g}_s(t,t') = \check{g}_{0}(t-t') + \check{g}_2 (t, t'), 
\end{equation}
where the Fourier transforms of $\check{g}_{0}$ and $\check{g}_{2}$ are defined as
\begin{align}
\check{g}_{0}(t-t') &= \int \frac{d\epsilon}{2\pi} e^{-i\epsilon(t-t')} \check{g}_{0}(\epsilon), \notag \\
\check{g}_2(t, t') &= \int \frac{d\epsilon}{2\pi} e^{-i\epsilon_+ t + i\epsilon_- t'} \check{g}_2(\epsilon_+, \epsilon_-) ,
\end{align}
with $\epsilon_\pm = \epsilon\pm \Omega$. Due to the normalization condition for the
stationary Green's function, $\check{g}_{0} \circ \check{g}_{0} = 1$, we obtain
$\check{g}_2 \circ \check{g}_{0} + \check{g}_{0} \circ \check{g}_2 = 0$ which becomes
$\check{g}_2(\epsilon_+,
\epsilon_-)\check{g}_{0}(\epsilon_-)+\check{g}_{0}(\epsilon_+)\check{g}_2(\epsilon_+,
\epsilon_-)=0$ in the energy domain.

	For the stationary quasiclassical Green's functions, we have \cite{noneqSC}
\begin{align}
& \hat{g}^{r(a)}_{0}(\epsilon) = g^{r(a)}_{0}(\epsilon)\hat{\tau}_3 + i\hat{\tau}_2
f^{r(a)}_{0}(\epsilon) , \\
& \hat{g}^{k}_{0}(\epsilon) = \left[\hat{g}^{r}_{0}(\epsilon) -
\hat{g}^{a}_{0}(\epsilon)\right] \tanh\left( \beta\epsilon/2\right) ,
\end{align} 
where $\beta = 1/(k_B T)$, and 
\begin{equation}
g_{0}^{r(a)}=f_{0}^{r(a)}\epsilon/\Delta_0 = \epsilon/s_0^{r(a)} ,
\end{equation}
with $s_0^{r(a)}=i\sqrt{\Delta_0^2-(\epsilon\pm i\gamma)^2}$. Here, $\gamma$ is the
phenomenological Dynes broadening parameter. Inserting Eqs.~\eqref{expand1_A} and
\eqref{expand2_A} into Usadel equation \eqref{Usadel_A} leads to
\begin{align}
& h_+ \check{g}_2(\epsilon_+,\epsilon_-) - \check{g}_2(\epsilon_+,\epsilon_-) h_- +
\check{\Delta}_2 \check{g}_{0}(\epsilon_-) - \check{g}_{0}(\epsilon_+) \check{\Delta}_2
\notag \\
& + iW_A \left[ \check{\bar{g}}_{0}(\epsilon)\check{g}_{0}(\epsilon_-) -
\check{g}_{0}(\epsilon_+)\check{\bar{g}}_{0}(\epsilon) \right] = 0 ,
\label{expand_A}
\end{align}
with $h_\pm = \epsilon_\pm \check{\tau}_3+\Delta_0 i\check{\tau}_2$, $W_A=D
e^2|\mathbf{A}_\Omega|^2$, and $\check{\bar{g}}_{0}= \check{\tau}_3 \check{g}_{0}
\check{\tau}_3$. By solving Eq.~\eqref{expand_A}, we obtain the retarded and advanced
components of the nonstationary term \cite{Heikkila19}, 
\begin{equation} \label{g_2A}
\hat{g}^{r(a)}_2(\epsilon_+,\epsilon_-) = \hat{g}^{r(a)}_{V}(\epsilon_+,\epsilon_-) +
\hat{g}^{r(a)}_{H}(\epsilon_+,\epsilon_-) ,
\end{equation}
with 
\begin{equation}
\hat{g}^{r(a)}_{V}(\epsilon_+,\epsilon_-) = iW_A \left[\frac{ \hat{\bar{g}}_{0}(\epsilon)
- \hat{g}_{0}(\epsilon_+) \hat{\bar{g}}_{0}(\epsilon) \hat{g}_{0}(\epsilon_-) }{s_+ + s_-}
\right]^{r(a)}  ,
\end{equation}
and 
\begin{equation} 
\hat{g}^{r(a)}_{H}(\epsilon_+,\epsilon_-) = i\Delta_2\left[\frac{ \hat{\tau}_2 -
\hat{g}_{0}(\epsilon_+) \hat{\tau}_2 \hat{g}_{0}(\epsilon_-) }{s_+ + s_-} \right]^{r(a)}.
\end{equation}
Here, 
\begin{equation*}
s_+^{r(a)} = i\sqrt{\Delta_0^2-(\epsilon_+\pm i\gamma)^2} , \ \ \
s_-^{r(a)} = i\sqrt{\Delta_0^2-(\epsilon_-\pm i\gamma)^2} ,
\end{equation*}
and $\hat{g}^{r(a)}_{V}$ and $\hat{g}^{r(a)}_{H}$ are due to the second order coupling to
the vector potential and the Higgs mode, respectively. 

The Keldysh component of $\hat{g}^k_2(\epsilon_+,\epsilon_-)$ can be expressed as the sum
of a regular part $\hat{g}^\mathrm{reg}_2$ and an anomalous part $\hat{g}^\mathrm{an}_2$
\cite{Efetov17,Artemenko79}, 
\begin{equation} \label{gk_2A}
\hat{g}^k_2(\epsilon_+,\epsilon_-) = \hat{g}^\mathrm{reg}_2(\epsilon_+,\epsilon_-) +
\hat{g}^\mathrm{an}_2(\epsilon_+,\epsilon_-),
\end{equation}
with
\begin{equation}
\hat{g}^\mathrm{reg}_2 = \hat{g}^r_2(\epsilon_+,\epsilon_-)\tanh(\beta\epsilon_-/2) -
\tanh(\beta\epsilon_+/2) \hat{g}^a_2(\epsilon_+,\epsilon_-) ,
\end{equation} 
and

\begin{equation}
\hat{g}^\mathrm{an}_2(\epsilon_+,\epsilon_-) =
\hat{g}^\mathrm{an}_V(\epsilon_+,\epsilon_-)+\hat{g}^\mathrm{an}_H(\epsilon_+,\epsilon_-).
\end{equation} 
The term $\hat{g}^\mathrm{an}_V$ is expressed as
\begin{align}
  \hat{g}^{\rm an}_V = & iW_A
\frac{[\tanh(\beta\epsilon_+/2)-\tanh(\beta\epsilon/2)][\hat{\bar{g}}^a_0 -\hat{g}^r_+
\hat{\bar{g}}^a_0 \hat{g}^a_-]}{s_+^r + s_-^a} \notag \\
+ & iW_A
\frac{[\tanh(\beta\epsilon/2)-\tanh(\beta\epsilon_-/2)][\hat{\bar{g}}^r_0
-\hat{g}^r_+ \hat{\bar{g}}^r_0 \hat{g}^a_-]}{s_+^r + s_-^a} ,
\end{align}
with the short notations $\hat{\bar{g}}^{r(a)}_0=\hat{\bar{g}}^{r(a)}_0(\epsilon)$ and
$\hat{g}^{r(a)}_\pm = \hat{g}^{r(a)}_0(\epsilon_\pm)$. The term $\hat{g}^\mathrm{an}_H$ is
given by
\begin{equation}
\hat{g}^\mathrm{an}_H =
[\tanh(\beta\epsilon_+/2)-\tanh(\beta\epsilon_-/2)]\hat{m}^\mathrm{an}_H ,
\end{equation}
where $\hat{m}^\mathrm{an}_H$ is obtained by replacing $\hat{g}^r_{0}(\epsilon_-)$ and
$s^r_-$ in the expression of $\hat{g}^r_H(\epsilon_+,\epsilon_-)$ with
$\hat{g}^a_{0}(\epsilon_-)$ and $s^a_-$, respectively.

From the superconducting gap equation \cite{noneqSC},
\begin{equation}
  \Delta(t)=-i\lambda \mathrm{Tr}[ \hat{\tau}_2 \hat{g}^k_s(t)], 
\end{equation}
where $\lambda$ parametrizes the pairing interaction, we have 
\begin{equation}
  \Delta_2 \int d\epsilon \ \mathrm{Tr} \left[ \hat{\tau}_2 \hat{g}^k_{0}(\epsilon) \right]
  = \Delta_0 \int d\epsilon \ \mathrm{Tr} \left[ \hat{\tau}_2
  \hat{g}^k_2(\epsilon_+,\epsilon_-) \right] .
\end{equation}
Lengthy but straightforward calculations lead to 
\begin{equation}
  \Delta_2 = iW_A \Delta_0\frac{B^r - B^a + B^\mathrm{an}}{C^r - C^a + C^\mathrm{an}},
\end{equation}
where 
\begin{align*}
B^{r(a)} &= 2 \int d\epsilon \ b^{r(a)}(\epsilon) \tanh(\beta\epsilon_\mp/2) , \\
B^\mathrm{an} & = 2 \int d\epsilon \ b_+^\mathrm{an}(\epsilon) \left[
\tanh(\beta\epsilon_+/2) - \tanh(\beta\epsilon/2) \right]  \\ 
              & + 2 \int d\epsilon \ b_-^\mathrm{an}(\epsilon) \left[
\tanh(\beta\epsilon/2) - \tanh(\beta\epsilon_-/2) \right], \\
C^{r(a)} &= \int d\epsilon \left[ c^{r(a)}(\epsilon) \tanh(\beta\epsilon_\mp/2) -
\tanh(\beta\epsilon/2)/s^{r(a)}_0 \right], \\
C^\mathrm{an} &= \int d\epsilon \ c^\mathrm{an}(\epsilon) \left[ \tanh(\beta\epsilon_+/2)
-\tanh(\beta\epsilon_-/2) \right] ,
\end{align*}
with 
\begin{align*}
b^{r(a)}(\epsilon) &= \left[\frac{\epsilon(\epsilon+\epsilon_+)s_- +
\epsilon(\epsilon+\epsilon_-)s_+}{s_+ s_0 s_- (s_+ + s_-)^2} \right]^{r(a)}, \\
b_+^{\rm an}(\epsilon) &= \frac{\epsilon(\epsilon+\epsilon_+)s_-^a +
\epsilon(\epsilon+\epsilon_-)s_+^r}{s_+^r s_0^a s_-^a (s_+^r + s_-^a)^2}, \\
b_-^{\rm an}(\epsilon) &= \frac{\epsilon(\epsilon+\epsilon_+)s_-^a +
\epsilon(\epsilon+\epsilon_-)s_+^r}{s_+^r s_0^r s_-^a (s_+^r + s_-^a)^2}, \\
\end{align*}
and
\begin{align*}
c^{r(a)}(\epsilon) &= \left[\frac{s_+s_- + \epsilon_+\epsilon_- + \Delta_0^2}{s_+ s_- (s_+
+ s_-)} \right]^{r(a)}, \\
c^{\rm an}(\epsilon) &= \frac{s_+^r s_-^a + \epsilon_+\epsilon_- + \Delta_0^2}{s_+^r
s_-^a (s_+^r + s_-^a)}.
\end{align*}
One can see that $c^\mathrm{an}$ is obtained from $c^r$ by replacing $s_-^r$ with
$s_-^a$.

\section{AC charge current} \label{appendixB}
The charge current $I(t)$ can be decomposed into a DC component $I_0$ and an AC component
$I_2$ with frequency $2\Omega$,  
\begin{equation}
I(t) = I_0 + I_2 e^{-i2\Omega t} .
\end{equation}
Using  circuit theory \cite{circuit99, Bruder02, circuit11}, the charge current can be
expressed as  
\begin{equation}
I(t) = \frac{2\pi}{4e} \mathrm{Tr}\big[ \hat{\tau}_3 \hat{I}^k(t) \big] ,
\end{equation}
with the matrix current
\begin{equation}
\hat{I}^k(t) = \frac{G_t}{2} \left[\check{g}_n, \check{g}_s \right]^k(t,t) ,
\end{equation}
where $G_t$ is the tunneling conductance of the NS junction. This leads to the following
expressions, 
\begin{align}
I_0 &= \frac{1}{4e} \int d\epsilon \ \mathrm{Tr} \big[ \hat{\tau}_3 \hat{I}^k_0(\epsilon) \big] , \\
I_2 &= \frac{1}{4e} \int d\epsilon \ \mathrm{Tr} \big[ \hat{\tau}_3 \hat{I}^k_2(\epsilon_+,\epsilon_-) \big] ,
\end{align}
with
\begin{align}
&\hat{I}^k_0(\epsilon) = \frac{G_t}{2}\big[ \hat{g}^r_n\hat{g}^k_0(\epsilon) +
  \hat{g}^k_n(\epsilon)\hat{g}^a_0 
-\hat{g}^r_{0} \hat{g}^k_n(\epsilon) -\hat{g}^k_{0}(\epsilon) \hat{g}^a_n \big]  , \notag
\\
&\hat{I}^k_2(\epsilon_+,\epsilon_-) = \frac{G_t}{2} \big[ \hat{g}^r_n
  \hat{g}^k_2(\epsilon_+,\epsilon_-) +
  \hat{g}^k_n(\epsilon_+)\hat{g}^a_2(\epsilon_+,\epsilon_-) \notag \\ 
&\qquad\qquad\qquad -\hat{g}^r_2(\epsilon_+,\epsilon_-) \hat{g}^k_n(\epsilon_-)
  -\hat{g}^k_2(\epsilon_+,\epsilon_-) \hat{g}^a_n \big] .
\end{align}

\begin{figure}
\centering
\includegraphics[width=\columnwidth]{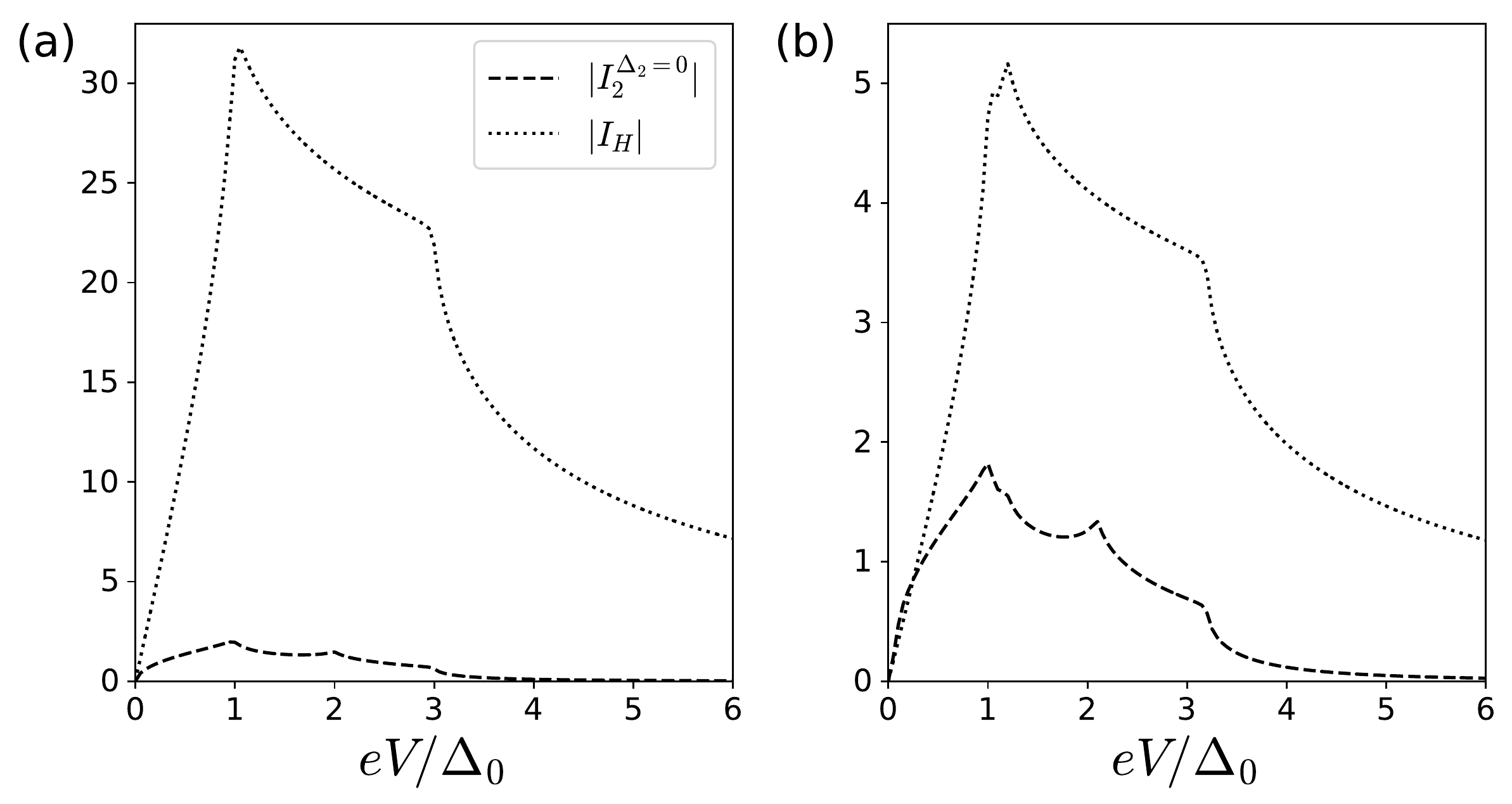} \\
\caption{The contributions from $\hat{g}_V^{r(a)}$ ($I_2^{\Delta_2=0}$, dashed lines) and $\hat{g}_H^{r(a)}$
  ($I_H$, dotted lines) in calculating $I_2$ using Eqs.~\eqref{I2_A} and \eqref{X} for (a)
$\Omega=\Delta_0$ and (b) $\Omega=1.1\Delta_0$. The currents are given in units of $G_t W_A/e$.
}
\label{fig_VD}
\end{figure}

Since $\mathrm{Tr} [\hat{g}^k_2(\epsilon_+,\epsilon_-)]=0$, the AC charge current can be
expressed as
\begin{equation} \label{I2_A}
I_2 = \frac{G_t}{8e} \int d\epsilon \ \mathrm{Tr} ( \hat{\tau}_3 \hat{X} ) ,
\end{equation}
with
\begin{equation} \label{X}
\hat{X} = \hat{g}^k_n(\epsilon_+)\hat{g}^a_2(\epsilon_+,\epsilon_-)
-\hat{g}^r_2(\epsilon_+,\epsilon_-) \hat{g}^k_n(\epsilon_-) . 
\end{equation}
From this expression, one can verify that the AC charge current vanishes in the absence of
the voltage bias, $I_2|_{eV=0}=0$. In Fig.~\ref{fig_VD}, we show the contributions from
$\hat{g}_V^{r(a)}$ and $\hat{g}_{H}^{r(a)}$ in $I_2$ using Eqs.~\eqref{I2_A} and
\eqref{X}.
The contributions to $I_2$ from $g_V$ (denoted as $I_2^{\Delta_2=0}$) and $g_H$ (denoted as
$I_H$) are calculated by replacing $g_2$ in Eq.~\eqref{X} with $g_V$ and $g_H$, respectively.
At resonance, i.e., $\Omega=\Delta_0$, the contribution from $\hat{g}_V^{r(a)}$ is much
smaller than that from $\hat{g}_{H}^{r(a)}$. This is due to the fact that $|\Delta_2|$
achieves its maximum and is much larger than $W_A$. For $\Omega=1.1\Delta_0$, the
contribution from $\hat{g}_V^{r(a)}$ cannot be ignored and can be even slightly larger
than that from $\hat{g}_H^{r(a)}$ at a small voltage bias.

\begin{figure}
  \centering
\includegraphics[width=\columnwidth]{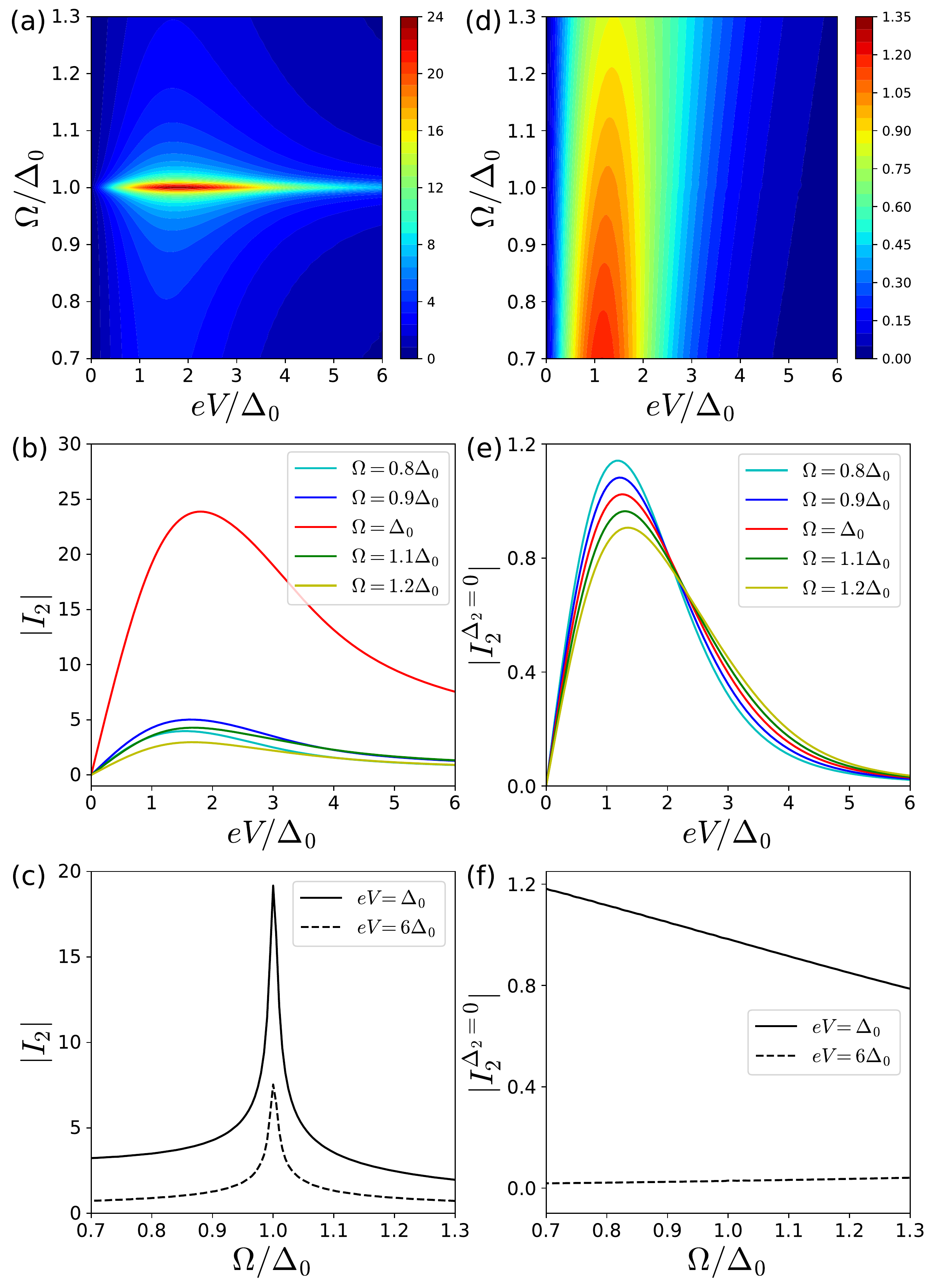} \\  
\caption{(a) AC charge current $|I_2|$ plotted against voltage bias $eV$ and irradiation
frequency $\Omega$. Panels (b) and (c) are line cuts of (a) at different $\Omega$ and
$eV$, respectively. (d)  AC charge current $|I_2^{\Delta_2=0}|$ in the absence of the
Higgs mode, i.e., by artificially setting $\Delta_2=0$. Panels (e) and (f) are line cuts
of (d) at different $\Omega$ and $eV$, respectively. The parameters are $T=0.7T_c$,
$\gamma=0.01\Delta_0|_{T=0}$, and $W_A=0.01\Delta_0|_{T=0}$. }
\label{fig_charge}
\end{figure}

We now try to get a compact expression for the AC charge current near resonance with
$\Omega\approx \Delta_0$ where the contribution from
$\hat{g}^{r(a)}_V(\epsilon_+,\epsilon_-)$ in Eq.~\eqref{g_2A} can be ignored.  Using
\begin{equation}
\hat{g}_{0}^r(\epsilon_+) \hat{\tau}_2 \hat{g}_{0}^r(\epsilon_-) = 2i\Delta_0\epsilon /
(s_+^r s_-^r) \hat{\tau}_3 + u \hat{\tau}_2 ,
\end{equation}
where the second term $u\hat{\tau}_2$ is unimportant in our derivation, Eq.~\eqref{I2_A}
can be expressed as 
\begin{align}
I_2 = \frac{G_t}{2e} \Delta_2 \Delta_0 \int d\epsilon \ \epsilon \bigg[
&\frac{\tanh(\frac{\epsilon_+-eV}{2k_BT})-\tanh(\frac{\epsilon_++eV}{2k_BT})}{s_+^a s_-^a
(s_+^a + s_-^a)}  \notag \\
- &\frac{\tanh(\frac{\epsilon_--eV}{2k_BT})-\tanh(\frac{\epsilon_-+eV}{2k_BT})}{s_+^r
s_-^r (s_+^r + s_-^r)} \bigg] .
\end{align}
Using the fact that $[s_+^{r(a)}]^2-[s_-^{r(a)}]^2=4(\epsilon \pm i\gamma)\Omega$, this
expression can be further simplified as,
\begin{equation}  
I_2 = \frac{G_t}{8e} \frac{\Delta_2}{\Omega} \int d\epsilon p(\epsilon) \Big[
\tanh(\frac{\epsilon -eV}{2k_BT})-\tanh(\frac{\epsilon +eV}{2k_BT}) \Big],
\end{equation}
with 
\begin{equation}
p(\epsilon) = \Delta_0 \bigg( \frac{1}{s_{++}^r} +
\frac{1}{s_{--}^a}-\frac{1}{s_0^a}-\frac{1}{s_0^r} \bigg),
\end{equation}
where 
\begin{align*}
s_{++}^{r} &= i\sqrt{\Delta_0^2-(\epsilon + 2\Omega + i\gamma)^2} , \notag \\
s_{--}^{a} &= i\sqrt{\Delta_0^2-(\epsilon - 2\Omega - i\gamma)^2} .
\end{align*}
Since $p(\epsilon)=p(-\epsilon)$, the expression for $I_2$ can be reduced to 
\begin{equation}
I_2 = \frac{G_t}{4e} \frac{\Delta_2}{\Omega} \int d\epsilon  p(\epsilon)
\tanh(\beta(\epsilon -eV)/2) ,
\end{equation}
which can also be written as 
\begin{equation}
I_2 = \frac{G_t}{4e} \frac{\Delta_2}{\Omega} \int d\epsilon q(\epsilon) \Big[
\tanh(\frac{\epsilon_- - eV}{2k_BT})-\tanh(\frac{\epsilon_+ - eV}{2k_BT}) \Big],
\end{equation}
with 
\begin{equation}
q(\epsilon) =  \Delta_0 \left( \frac{1}{s_{+}^r} - \frac{1}{s_{-}^a} \right). 
\end{equation}
Since $q(\epsilon) \approx 2\Delta_0/\epsilon$ for $\epsilon \gg \Delta_0$, we have $I_2
\propto \Delta_2/(eV)$ at large bias with $eV\gg \Delta_0$. In other words, at resonance
and for large voltage bias, the AC current is inversely proportional to the voltage bias.

Similarly to Fig.~(2) in the main text, we show the AC charge current behavior at a large
temperature with $T=0.7T_c$ in Fig.~\ref{fig_charge}. A resonant behavior can be still
observed at $\Omega=\Delta_0$. Fig.~\ref{fig_charge}(b) also shows that $|I_2|$ exhibits a
non-monotonic behavior with increasing bias and vanishes inversely proportional to the
bias. Due to thermal excitations, the maximum of $|I_2|$ is shifted to larger voltages
with $eV > \Delta_0$ compared to Fig.~(2) in the main text. Kinks and local maxima are
smeared by the thermal excitation as well.

\end{document}